\documentclass[nofootinbib,prd,twocolumn]{revtex4}
\usepackage{amstext,amsmath,amssymb,amsfonts,bbm}
\usepackage{amsthm}
\usepackage[dvips]{graphicx}
\usepackage{euscript}
\usepackage{epsfig}

\def\be{\begin{equation}}
\def\ee{\end{equation}}
\def\bes{\begin{eqnarray}}
\def\ees{\end{eqnarray}}
\def\nn{\nonumber}
\def\arr{\rightarrow}
\def\6{\langle}
\def\9{\rangle}
\def\half{\mbox{$\f 1 2$}{}}

\def\f{\frac}

\DeclareMathOperator{\tr}{tr}

\def\1{{\mathbbm 1}}

\def\com{{\mathbb{C}}}

\def\he{\hat{E}}

\def\ch{{\cal H}}
\def\ci{{\cal I}}

\def\jj{{\mathbf{J}}}
\def\hjj{{\hat{\mathbf{J}}}}
\def\hj{{\hat{J}}}

\def\hll{{\hat{\mathbf{L}}}}

\def\hmm{\hat{{\mathbf{M}}}}
\def\hm{\hat{M}}

\def\bn{\mathbf{n}}

\def\hu{{\hat{U}}}
\def\hx{{\hat{X}}}
\def\inr{{\mathrm{in}}}

\begin{document}
\title{Quantum tetrahedron and its classical limit}

\author{ Daniel R. Terno}\email{dterno@physics.mq.edu.au}
\affiliation{Centre for Quantum Computer Technology,
Department of Physics, Macquarie University, Sydney NSW 2109,
Australia}

\begin{abstract}Classical information that is retrieved from a quantum tetrahedron
is intrinsically fuzzy. We present an asymptotically optimal
generalized measurement for the extraction
 of classical information from a quantum tetrahedron. For a single tetrahedron the optimal
 uncertainty in dihedral angles is shown to scale as an inverse of the surface
 area. Having  commutative observables allows to show  how the clustering of many small tetrahedra leads to a
 faster convergence to a classical
 geometry.
\end{abstract}

\maketitle


\section{Introduction}
 Basis states in the
kinematical Hilbert space of loop quantum gravity (LQG) are
represented by spin networks, which are  finite,  directed, labeled
graphs. Their edges are labeled by SU(2) spins, and SU(2)-invariant
tensors (intertwiners) label the  vertices. These two families of
decorations  are linked to the geometric operators. A spin label
attached to a link determines the area of  a surface that is
intersected by  it, while an intertwiner is associated with the
volume of a spatial region that contains the vertex.   One of the
major results of LQG \cite{lqg} is a quantization of space: the
spectrum of geometric operators representing area or volume is
discrete. At present there is no rigorous proof that this result is
valid in the full theory, since the geometric operators do not
commute with the constraints, but there are plausible arguments that
this is indeed the case.
\cite{contr}.

The simplest state that represents a finite volume of space involves
a four-valent vertex spin network vertex.   A dual picture
associates vertices with tetrahedra and edges with two-dimensional
surfaces, so an atom of space can be represented as a quantum
tetrahedron. Its triangular areas fix the spin labels and the volume
determines the intertwiner. The same mathematical structure results
from a formal quantization of a classical tetrahedron
\cite{tetra}.

Geometry of a classical tetrahedron is determined by  six
parameters. This number
 is  obtained by removing  the rigid body degrees of freedom
from  the twelve coordinates of tetrahedron's vertices. Six
geometric variables are associated with observables on the Hilbert
space $\ch_0$ of a quantized tetrahedron. However, the maximal set
of commuting observables contains only five operators,  which makes
a retrieved classical description intrinsically fuzzy.  A coherent
quantum
\cite{cs} tetrahedron is a useful toy model to investigate a
classical limit of LQG \cite{lqg}. Recently it gained  importance in
providing boundary states in the studies of a graviton propagator
\cite{prop}.

In the language of quantum information the classical limit of a
quantum tetrahedron is equivalent to a faithful transmission of its
shape without any knowledge of the spatial orientation. We will use
quantum-informational techniques to get some insights into the
classical limit.

 This paper is organized as follows. In the next Section we review
 some properties of classical and quantum tetrahedra. Sec.~III
 establishes an upper bound on the convergence to a sharp classical
 geometry. Sec.~IV exhibits a  generalized measurement
 of  two  parameters that correspond to the non-commuting
 observables. Finally, Sec.~V discusses
 classical geometry of an aggregate of many small tetrahedra.
 Necessary background information about generalized measurements is
 given in Appendix A, while the explicit formulas are collected in
 Appendix B.

\section{Classical and quantum tetrahedra }
In this Section we review some facts about classical  and quantum
\cite{tetra,ct} tetrahedra. There are several sets of  six numbers that
determine its shape.
  For example, one can use
 the six edges of a tetrahedron, or four facial areas and two independent angles between them. The latter
  parametrization gives a natural way to compare  classical values with the estimates obtained from a quantum
 tetrahedron.

We label the outward normals to the faces as $\jj_i$, and following
the convention take their lengths to be twice the triangular areas,
$J_i=2A_i$. Being a closed surface
 a tetrahedron satisfies the closure condition
\be
\sum_{i=1}^4\jj_i=0. \label{closure}
\ee
There is a number of useful relations between areas, angles and
volume \cite{ct}. Angles between the triangular faces are the
(inner) dihedral angles, which are related to the  outer dihedral
angles $\theta_{ij}$,
\be
J_{ij}\equiv\jj_i\!\cdot\jj_j=J_iJ_j\cos\theta_{ij} \label{angle}
\ee
as $\theta_{ij}^\inr=\pi-\theta_{ij}$. The volume can be expressed in terms of the area vectors,
\be
V^2=-\f{1}{36}\epsilon_{abc}J_1^a J_2^b J_3^c=-\f{1}{36}\jj_1
\cdot\jj_2\times\jj_3.
\ee
 and, e.g.,
\be
\sin\theta_{ij}^\inr=\f{3V\, l_{ij}}{2 A_i A_j},
\ee
where $l_{ij}$ is a length of the edge between the faces $i$ and
$j$. In the following we take $J_1$,$J_2$, $J_3$, $J_4$,
$\theta_{12}$ and $\theta_{23}$ to form  the shape-defining set.

In the quantized problem the four normals are identified with the
generators of SU(2),
\be
\hjj^2_i|j_i,m_i\9=j_i(j_i+1)|j_i,m_i\9,\qquad
\hat{J}_{zi}|j_i,m_i\9=m_i|j_i,m_i\9.
\ee
For  later use we note that the Casimir operator asymptotically
equals to $J\equiv\sqrt{j(j+1)}\simeq j+\half$.

The closure constraint Eq.~(\ref{closure}) restricts the Hilbert
space to its SU(2) invariant subspace,
\be
\ch_0=\bigoplus_{\{j_k\}}\ch_0^{\{j_k\}}=\bigoplus_{\{j_k\}}
\mathrm{Inv}\left(\bigotimes_{k=1}^4\mathbbm{C}^{d_k}\right),
\ee
where the dimension of a spin-$j_k$ representation is $d_k\equiv
d_{j_k}=2j_k+1$. The states on $\ch_0$ are identified with the
intertwining maps
\be
\ci_{\{j_k\}}:\bigotimes_{k=1}^4\mathbbm{C}^{d_k}\arr\mathbbm{C}.
\ee
The basis can be  constructed in two steps,  e.g., first by coupling
spins 1 with 2,
\be
\hll_{12}=\hjj_1+\hjj_2,\qquad \hat{J}_{12}=\hjj_1\cdot\hjj_2,
\ee
and 3 with 4, and then forming  the singlets from the intermediate
pairs
\be
|l\9=\sum_m\f{(-1)^{l-m}}{\sqrt{2j+1}}|l,m\9_{12}|l,-m\9_{34},
\ee
where we suppressed the labels $j_1,\ldots, j_4$. The singlets are eigenvalues of $\hll_{12}^2$.
An alternative
basis is defined by the eigenvalues of $\hll_{23}^2$. The two sets
of basis vectors are related through 6$j$-symbols,
\be
\6l|l'\9=(-1)^{\sum j_i}\sqrt{d_ld_{l'}}
\left\{\begin{array}{ccc} j_1 & j_2 & l \\j_3 & j_4 &
l'\end{array}\right\}.
\ee
The commutator of $\hj_{12}$ and $\hj_{23}$ is related to the volume
through \cite{le}
\be
[\hj_{12},\hj_{23}]=-i\epsilon_{abc}\hj_1^a \hj_2^b \hj_3^c\equiv
i\hu.
\ee
In agreement with  LQG results its absolute value operator $|\hu|$
is identified with the quantized classical squared volume $36V^2$.

There are no self-adjoint angle operators $\hat{\theta}$  on
finite-dimensional representations of SU(2) \cite{hol}, so a
classical angular parameter $\theta$ is estimated by using a
positive operator-valued measure (POVM). General properties of such
measures are given in  Appendix A. Following
\cite{tetra,cs} we restrict ourselves to the case where four areas
have well-defined values. Then the angles are easily identified
through the quantum analog of Eq.~(\ref{angle}), where the
operators $\hat{J}_{ij}$ are extracted from
\be
\hll_{ik}^2=\hjj_i^2+\hjj_k^2+2\hj_{ik}.
\ee

 To simplify the following formulas we consider the case of four
 equal areas: $j_1=\ldots=j_4=j$. Accordingly,
the estimate of the classical angle $\theta_{12}$ is given by
\be
z_{ik}(\rho)\equiv\cos\theta_{ik}(\rho)=\frac{\6\hj_{ik}\9}{j(j+1)}.
\label{z}
\ee

An obvious spread estimator of a classical random variable $z$ with
a probability distribution $p(z)$, $\int p(z)dz=1$,  is the standard
deviation
\be
(\Delta z)^2\equiv \Delta^2 z=\6z^2\9-\6z\9^2,
\ee
where the statistical moments are calculated  with respect to $p(z)$.

In quantum theory a pair of non-commutative self-adjoint operators
satisfies the Schr\"{o}dinger-Robertson relation \cite{hol,bgl}. In
the case of $\hj_{12}$ and $\hj_{23}$ it is
\be
\Delta^2(\hj_{12})\Delta^2(\hj_{23})\geq
{\small\mbox{$\f{1}{4}$}}|\6\hu\9|^2+\sigma^2(\hj_{12},\hj_{23}),
\ee
where
\be
\Delta^2\hx \equiv\6\hx^2\9-\6\hx\9^2,
\ee
and
\be
\sigma(\hx,\hat{Y})
\equiv\6\hx\hat{Y}+\hat{Y}\hx\9/2-\6\hx\9\6\hat{Y}\9.
\ee

A good semiclassical state that describes a tetrahedron with fixed
triangular areas should have small angular and volume uncertainties,
\be
\f{\Delta z_{ik}}{\6z_{ik}\9}\arr 0, \qquad \f{\Delta V}{\6V\9}\arr 0
\ee
across the range of the angle and volume expectations that
correspond to the classical values.

When Eq.~(\ref{z}) is used to obtain the dihedral angles, then is
easy to see that for such a state
\be
\Delta z_{12}\Delta
z_{23}=\f{\Delta(\hj_{12})}{J^2}\f{\Delta(\hj_{23})}{J^2}\geq\f{|\6\hu\9|}{2
J^4}\sim\f{1}{J} \label{hei}
\ee
Rovelli and Speziale \cite{cs} constructed a family of such states.
We discuss them in Sec.~\ref{joint}.

\section{The optimal convergence rate}

 Minimization of $\Delta z_{12}\Delta z_{23}$  does not necessarily produce the best semiclassical
 states. For example, the eigenstates of either of $\hj_{ik}$
 result in $\Delta z_{12}\Delta z_{23}\equiv 0$.
 Moreover, fixing  the  expectation values $z_{12}$ and $z_{23}$ does not
fix the expectation of $\hu$. The volume can be recovered by
classical calculation from the determined values of $\theta_{ik}$,
but the physical significance of the states with $36V^2_{\rm{
class}}\neq\6\hu\9$ is not clear.

To study the asymptotics let us introduce some rescaled quantities.
We set
 $\6\hj_{12}\9=\ell_*K^2$, $\6\hj_{23}\9=k_*K^2$ and $J=j_*K$. Thus the
 shape
\be
z_{12}=\frac{\ell_*}{j_*}, \qquad z_{23}=\frac{k_*}{j_*},
\ee
 is fixed even if the size goes to infinity, $K\rightarrow \infty$.

If the goal is to minimize the left hand side of Eq.~(\ref{hei}),
then not fixing the volume expectation leaves a larger parameter
space, and one can expect a smaller product of variances.
 Appearance of the  eigenvalues of $\hu$ in pairs $\pm u$ makes it appealing
 to expect the optimal states to have a zero expectation of the
 squared volume. It is so, e.g, in $j=1$ case. Then a detailed analysis
 shows that for all expectations $\6\hj_{12}\9=\ell_0$, $\6\hj_{23}\9=k_0$,
 the minimal value of $\Delta^2(\hj_{12})\Delta^2(\hj_{23})$ is
 reached on the states that have $\6\hu\9=0$.

Nevertheless, even the unconstrained minimization gives $\Delta
z_{12}\Delta z_{23}\propto1/K$. This rest of this Section deals with
derivation of this bound. We introduce another parametrization of a
tetrahedron which
 allows to map the unconstrained search of the minimal
uncertainty states to the problem of optimal direction transmission
\cite{hol,str}.

The four vectors of equal length that satisfy the closure condition
(\ref{closure}) can be represented as
\begin{align}
\jj_{1,2}=&\,J(\mp\sin\theta,0,\cos \theta), \nn \\
\jj_{3,4}=&\,J(\pm\sin\theta\cos\phi,
\pm\sin\theta\cos\phi,-\cos\theta). \label{repangle}
\end{align}
Using the  relations for the dihedral angles we find that
\begin{align}
2\theta = &\,\theta_{12}=\pi-\theta_{12}^\inr \nn, \\
\cos\phi = &\, \f{\cos\theta_{23}+\cos\theta}{\sin^2\theta}.
\end{align}
Since
\be
V^2=-\f{1}{18}J^3\cos\theta\sin^2\theta\sin\phi,
\ee
the admissible range of the parameters is
\be
0\leq\theta\leq \pi/2, \qquad \pi/2\leq\phi\leq 3\pi/2.
\ee
For example, a regular tetrahedron is parameterized by
$\cos\theta=1/\sqrt{3}$  and $\phi=\pi/2$.

 Eq.~(\ref{repangle})
 maps a problem of finding the minimum of $\Delta
z_{12}\Delta z_{23}$ to  the task of finding the optimal direction
transmission protocol. Given a spatial direction
$\bn_0(\theta_0,\phi_0)$ there is an encoding and decoding scheme
that results in the optimal estimate $\bn(\theta,\phi)$.  The
protocol is optimal with respect to the \textit{fidelity} and
related error measures \cite{hol,str}. Fidelity is defined as
\be
F=\f{1+\6\cos\chi\9}{2},
\ee
where $\cos\chi=\bn_0\cdot\bn$, and the average is taken over the
resulting probability distribution $p(\theta_0,\phi_0;\theta,\phi)$.
SU(2) covariance properties allow to reduce the problem to sending a
single fiducial direction. A complementary quantity $D=1-F$  is the
mean square error of the measurement, if the error is defined as
$\sin^2\chi/2=|\bn_0-\bn|^2$.

 There are many different versions of this communication task. They
 differ in   physical systems that serve as information carriers  (see \cite{hol,str}
 and the references therein).
  However, in all of them the optimal
 measurement is unbiased, i.e.,
\be
\6\theta\9=\theta_0,\qquad \6\phi\9=\phi_0, \label{nobias}
\ee
and the fidelity asymptotically approaches unity as
\be
D\propto\f{1}{d},\qquad d\arr\infty,
\ee
where $d=\dim\,\ch$ is the dimension of the available Hilbert space,
and a pre-factor depends on the set-up details. If the information
carrier is a single spin-$j$ particle, then $d=d_j=2j+1$. On the
other hand, for a system that consists of $N$ two-level systems
(qubits)  the total Hilbert space is
\be
\left.\com^{2}\right.^{\otimes N}=\sum_jV^j\otimes\com^{d_j},
\ee
where each term in the sum is a direct product of an appropriate
degeneracy space and spin-$j$ irreducible representation, but the
available space is much smaller. Only a single copy of each of the
representation spaces can be used for the direction transmission,
and $d=\sum_jd_j\simeq N^2/4$.

In our case the Hilbert space is the
intertwiner space, and the assumption $j_1=\ldots=\l_4=j$ gives
\be
d=2d_j.
\ee
To compare this result with the product of variances $\Delta
z_{12}\Delta z_{23}$ we first express the deviation angle $\chi$ in
terms of $\delta\theta=\theta-\theta_0$ and
$\delta\phi=\phi-\phi_0$. Assuming  $|\delta\theta|<\theta_0$,  it
is easy to derive the asymptotic result
\be
\chi^2\simeq\sin^2\theta_0\delta\phi^2+\delta\theta^2.
\ee
From Eq.~(\ref{nobias}) it follows that
\begin{align}
\Delta^2(z_{12})& \simeq
4\sin^2(2\theta_0)\6\delta\theta^2\9, \\
\Delta^2(z_{23})&\simeq\sin^4(\theta_0\!/2)\sin^2\phi_0\6\delta\phi^2\9 \nonumber \\ &+
4\sin^2(2\theta_0)\cos^4(\phi_0\!/2)\6\delta\theta^2\9,
\end{align}
and
\begin{align}
\Delta^2(z_{12})\Delta^2(z_{23})&\simeq
16\cos^2\theta_0\sin^6\theta_0\sin^2\phi_0\6\delta\theta^2\9\6\delta\phi^2\9\nonumber
\\
&\leq 4\cos^2\theta_0\sin^4\theta_0\sin^2\phi_0\chi^4.
\label{delta2}
\end{align}
Taking into account that $ D\simeq\chi^2/4$, we see that the
uncertainty of the optimal shape transmission
\be
\Delta(z_{12})\Delta(z_{23})=c(k_*,l_*)D\propto\f{1}{d}=\f{1}{2j_*K},\label{upper}
\ee
where $c(k_*,l_*)$ follows from Eq.~(\ref{delta2}). Hence even if we disregard the physical meaning
of the resulting states, the uncertainty still behaves as $1/J$.

\section{Joint POVM and its optimality} \label{joint}
Expressions like Eq.~(\ref{hei}) refer to an ensemble of identically
prepared systems,  where  in half of the cases one measures
$\hj_{12}$ and in the other half $\hj_{23}$. Since  classical
dihedral angle variables $J_{12}$ and $J_{23}$ are defined
simultaneously, the emergence of classicality is properly described
\cite{hol, bgl} by  convergence of the joint probability distribution
$p(J_{12},J_{23})$  to their sharp classical profiles. This is
achieved by a POVM that is described in this Section.

 A generic pure state on   $\ch_0^{\{j_k\}}$
 is given by
\be
|\psi\9=\sum_{l=0}^{2j}c_l|l\9, \qquad c_l\equiv
|c_l|e^{i\phi_l}.\label{state}
\ee
For example, if the sextet of the classical parameters of a
tetrahedron is completed by
\be
j_{12}\equiv\ell_0=\6\hj_{12}\9,\qquad k_0=\6\hj_{23}\9,
\ee
then the amplitude of the states proposed in
\cite{cs} has a Gaussian profile centered around $l_0$,
$j_{12}=\half l_0(l+1_0)-\half j_1(j_1+1)-\half j_2(j_2+1)$.

 The phase is determined
with the help of an auxiliary tetrahedron that  is made from the
four edges $j_1+\half,\dots,l_4+\half$ and $j_0+\half$ and
$k_0+\half$. It  equals to
\be
\phi_l=\phi_0 l,
\ee
where $\phi_0$ is dihedral angle between the two faces of an
auxiliary tetrahedron that share a link $l_0+\half$.

If a state $|\psi\9$ is  is sharply peaked on some $l_0$, a useful
asymptotic expression for the volume can be derived as follows.
For any state
\begin{align}
\6\psi|\hu|\psi\9&=\sum_l
c_{l+1}^*c_l\hu_{l+1,l}+c_l^*c_{l+1}\hu_{l,l+1}\nonumber \\ &=2
\mathrm{Re}\!\left(\sum_l c_{l+1}^*c_l\hu_{l+1,l}\right).
\end{align}
Hence the leading order  asymptotic expression is
\begin{align}
\6\psi|\hu|\psi\9&\simeq 2\mathrm{Re}\sum_l
|c_l|^2 \exp[-i(\phi_{l+1}-\phi_{l})]\hu_{l+1,l}\nonumber \\ &\simeq
2\mathrm{Re}\!\left(
\exp[-i(\phi_{l_0+1}-\phi_{l_0})]\hu_{l_0+1,l_0}\right)\nonumber \\ &=2\sin(\phi_{l_0+1}-\phi_{l_0})\hu_{l_0+1,l_0}.
\end{align}

Partial traces of an arbitrary $|\psi\9\in\ch_0^{\{j\}}$ on
$\ch_{12}=\mathbbm{C}^{d_1}\otimes\mathbbm{C}^{d_2}$
\be
\rho_{12}=\sum_{l}\frac{|c_l|^2}{2l+1}\sum_{m=-l}^l|l,m\9\6l,m|,
\label{rho}
\ee
and on $\ch_{23}$
\be
\rho_{23}=\sum_{l'}\frac{|k_{l'}|^2}{2l'+1}\sum_{m=-l'}^{l'}|l',m\9\6l',m|.
\ee
are diagonal, a feature we use below.

To construct a joint measurement that results in $z_{12}$ and
$z_{23}$ we use a ``commutative spin observable".  A POVM that is used to
identify the directions is built from the normalized SU(2) coherent states,
\be
\hat{E}_{\theta\phi}=\f{2j+1}{4\pi}|\theta,\phi\9\6\theta,\phi|,
\qquad \int\hat{E}_{\theta\phi}d\Omega_{\theta\phi}=\1,
\ee
where
\begin{align}
|\theta,\phi\9&=\sum_{m=-j}^j  D^{\,(j)}(\phi, \theta,\psi=
0)_{mj}|j,m\9
\nonumber\\
&=\sum_{m=-j}^j{{2j}\choose{j+m}}^{1/2}\cos^{j+m}\theta\!/2\sin^{j-m}\theta\!/2
\,e^{-im\phi}|jm\9.
\end{align}
A commutative angular momentum observable is a collection of (statistical) moment operators that correspond
to this POVM.
The first moment operator is used to calculate the expectation values,
\be
\hmm^{(1)}=(j+1)\int \mathbf{n}\hat{E}_{\theta\phi}d\Omega_{\theta\phi},
\ee
where  $\mathbf{n}=
(\sin\theta\cos\phi,\sin\theta\sin\phi,\cos\theta)$ is a  unit
vector. The measurement is  unbiased, i. e., for any state $\rho$
\be
\tr (\hmm^{(1)}\rho)=\tr(\hjj\rho),
\ee
but the expectation of the second moment operator is never zero.

Using $\cos\theta_{12}=\bn_1\cdot\bn_2$, an unsharp measurement of $J_{12}$ can be described by a POVM
\begin{widetext}
\be
\he_{z_{12}}dz_{12}=dz_{12}\int\delta(\cos\theta_1\cos\theta_2-\cos(\phi_1-\phi_2)\sin\theta_1\sin\theta_2-z_{12})
\hat{E}_{\theta_1\phi_1}\hat{E}_{\theta_2\phi_2}d\Omega_1
d\Omega_2.
\ee
\end{widetext}
This expression is hard to manipulate and the statistical moments
for $\jj_{ik}$ are
 obtained through the integration over the original angles,
\be
\hm^{(1)}_{ik}=(j+1)^2\int (\bn_1\!\cdot\!\bn_2)\hat{E}_{\theta_i\phi_i}\hat{E}_{\theta_k\phi_k}d\Omega_i
d\Omega_k \label{firstm}
\ee
and
\be
\hm^{(2)}_{ik}=(j+1)^4\int (\bn_1\!\cdot\!\bn_2)^2\hat{E}_{\theta_i\phi_i}\hat{E}_{\theta_k\phi_k}d\Omega_i
d\Omega_k.
\ee
Their matrix elements are spelled out in Appendix B. These operators
have a simple overall structure. In particular, in the corresponding
$(l,m)$ basis they have a form
\be
\hm^{(i)}_{lm,l'm'}=\hm^{(i)}_{lm,l'm}\delta_{mm'}.
\ee
Unlike the sharp projective measurements of $\jj_{12}$ and $\jj_{23}$, the above construction allows
 a simultaneous estimate of the  angles $\theta_{12}$ and $\theta_{23}$.

 It
follows  from Eq.~(\ref{rho}) that we are interested in the
asymptotic behavior of their expectation values on the states
\be
\rho_l=\f{\1_l}{2l+1},
\ee
where  $j=j_*K$, $l=l_*K$, and $K$ goes to infinity. The operator
$\hm^{(1)}_{12}$ is not an unbiased estimator: for a generic
$|\psi\9\in\ch_{12}$ the  expectation $\6\psi|\hm^{(1)}_{12}|\psi\9$
is different from $\6\psi|\hj_{12}|\psi\9$. However, it is possible
to show that
\be
\tr(\hm^{(1)}_{12}\rho_l)=\tr(\hj_{12}\rho_l)=j_{12}=\half l(l+1)-j(j+1).
\ee
The asymptotics of $\Delta z_{12}$ was investigated both
analytically and numerically. It was found that for $\rho_l$
\be
\lim_{K\arr\infty}\frac{\sigma^2_{j_{12}}(M)}{\6J_{12}\9^2}
\equiv\f{\6\hm^{(2)}_{12}\9-j_{12}^2}{j_{12}^2}\propto\f{1}{K}.\label{asymK}
\ee
In particular, if $l=0$ then $j_{12}=-j(j+1)$ and
\be
\f{\6\hm^{(2)}\9_{\rho_0}-j_{12}^2}{j_{12}^2}=\f{2j+1}{j(2j+3)}\simeq
\f{1}{j}.
\ee

A simple error analysis  shows that if the state is peaked on
$\6\psi|\hj_{12}|\psi\9\simeq j_{12}  $ the variance of the
unsharply measured $J_{12}$ is the sum of the sharp variance
$\Delta^2(\hj_{12})=\6\psi|\hj^2_{12}|\psi\9-\6\psi|\hj_{12}|\psi\9^2$
and the measurement unsharpness
$\6j_{12}|\hm^{(2)}_{12}|j_{12}\9^2$,
\be
\Delta^2(J_{12})\simeq\Delta^2(\hj_{12})+\sigma^2_{j_{12}}(M).
\ee
 Hence
the result of Eq.~(\ref{asymK}) guaranties that if a state is such
that $\6\hj_{12}\9=\ell_*K^2$ , $\6\hj_{23}\9=k_*K^2$ and
\be
\Delta(\hj_{12})\Delta(\hj_{23})\sim K^3,
\ee
as the states of \cite{cs} are, then the estimate obtained from the joint POVM  asymptotically
behaves as
\be
\Delta z_{12}\Delta z_{23}\sim 1/K,
\ee

\section{Fast convergence to the classical limit}
While we established that the uncertainties in the shape of an atom
of space scale inversely with its surface area, it is interesting to
investigate  convergence of  geometry to its classical value for
more complicated structures. Our goal is to check the intuitive
assumption that ``many small tetrahedra approach the classicality
faster than just a scaled up single tetrahedron".

First we subdivide a single classical tetrahedron through a number
of iterative steps that are described below. \textit{Subdivision} is
an extensively studied technique in computer aided geometric design
and visualisation, as well as in numerical analysis, particularly in
computational fluid dynamics  (see, e.g.,
\cite{subtet,subtet2} and references therein).
Given this refined triangulation we set up a refined spin network  as a dual of the new
triangulation, and label its edges according to the triangular areas
they pierce.

\begin{figure}[htbp]
    \begin{minipage}{\columnwidth}
    \begin{center}
        \resizebox{0.7\textwidth}{!}{\includegraphics{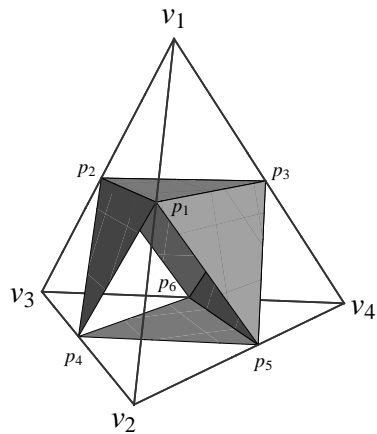}}
    \end{center}
    \end{minipage}
    \caption{\small{
The first stage of the regular subdivision: four child tetrahedra
and an octahedron. The faces $(p_1p_4p_5)$ and $(p_2p_4p_6)$ are
transparent.}}\label{choiplot}\vspace{-1mm}
\end{figure}

At every step the most direct approach results in dividing a
tetrahedron into eight descendants. The four tetrahedra are obtained
by cutting off the corners of the parent tetrahedron at the edge
midpoints, as shown on Fig~1. They are obviously similar to their
parent.  Each of its faces is now composed of the outer faces of
three child tetrahedra and one of the faces of an octahedron. The
remaining octahedron is split into two pyramids, each of which is
separated into two tetrahedra. This splitting depends on the choice
of the interior diagonal, so there are three possibilities for this
subdivision. In any case, the resulting tetrahedra are not similar
to the parent one. There are at least three different similarity
classes for the tetrahedra. Moreover,  for generic initial
tetrahedra a compliance with naturally defined requirements of
nestedness, consistency and stability of the subdivisions is not
guarantied
\cite{subtet2}.

We use this scheme only at the last iteration, to produce a
four-valent spin network. In all other step w use a different
subdivision scheme for octahedra. This refinement rule consists in
subdividing an octahedron into six child octahedra and eight
tetrahedra by connecting the edge midpoints of each face (Fig.~2(a))
and by connecting all edge midpoints to the barycenter of the parent
octahedron (Figs.~2, 3). Even for an arbitrary initial tetrahedron
its barycenter
\be
b=\frac{1}{4}(v_1+v_2+v_3+v_4),
\ee
coincides with the barycenter of the child octahedron, which ensures
that the eighth second generation tetrahedra are similar to the
initial one with the scale factor $1/4=(1/2)^2$.
\begin{figure}[htbp]
\epsfxsize=0.3\textwidth
 \epsffile{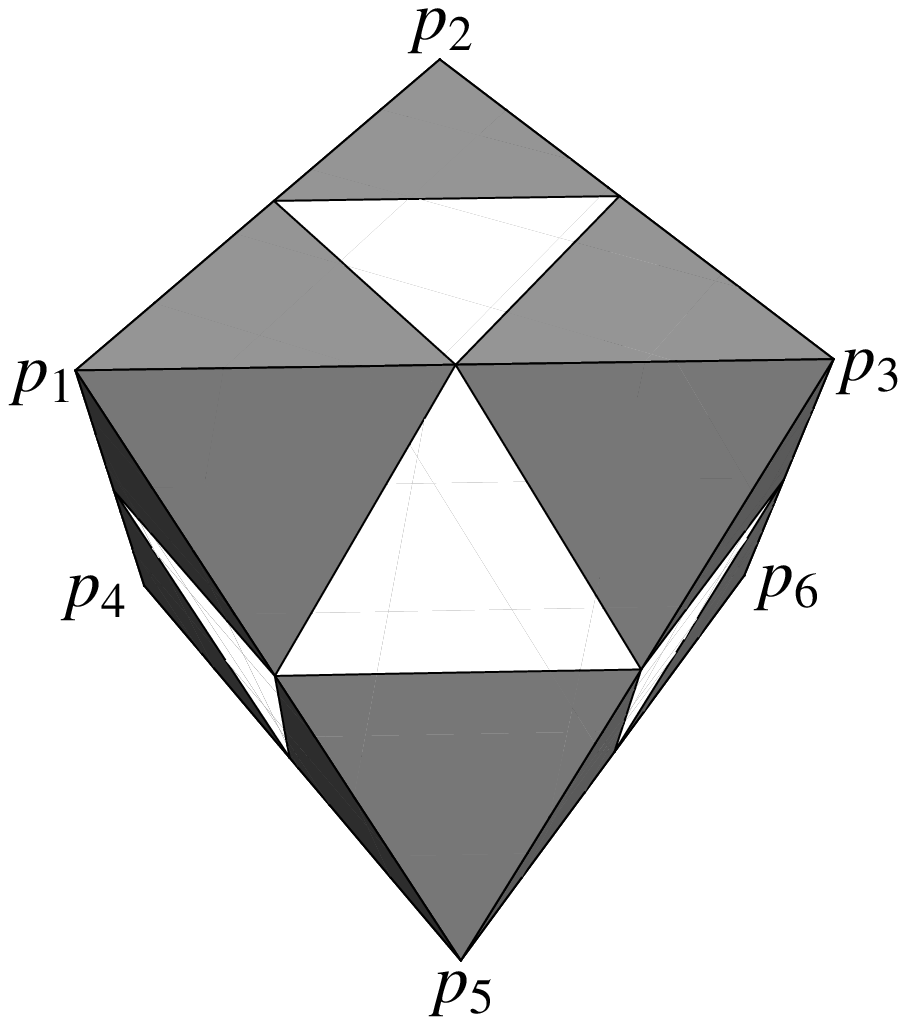} \hspace{0.4cm}{ \epsfxsize=0.3\textwidth\epsffile{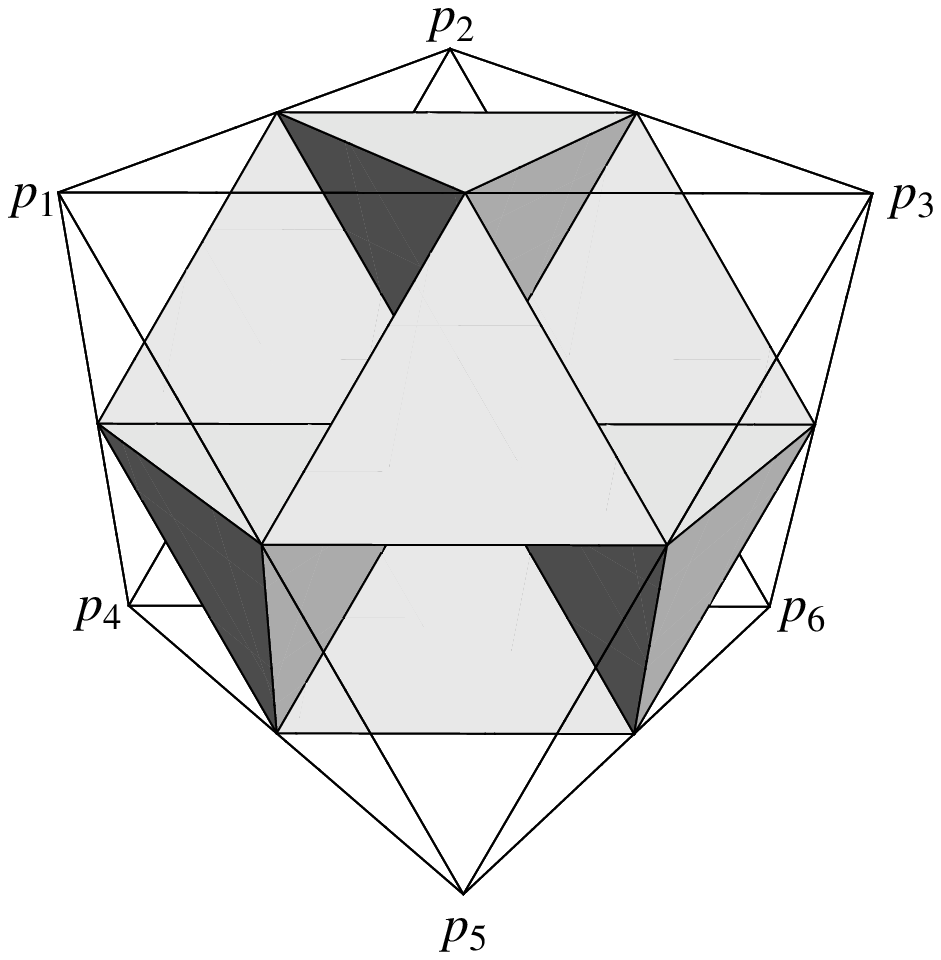}}
 \hspace{0.4cm} \raisebox{0.6cm} { \epsfxsize=0.22\textwidth\epsffile{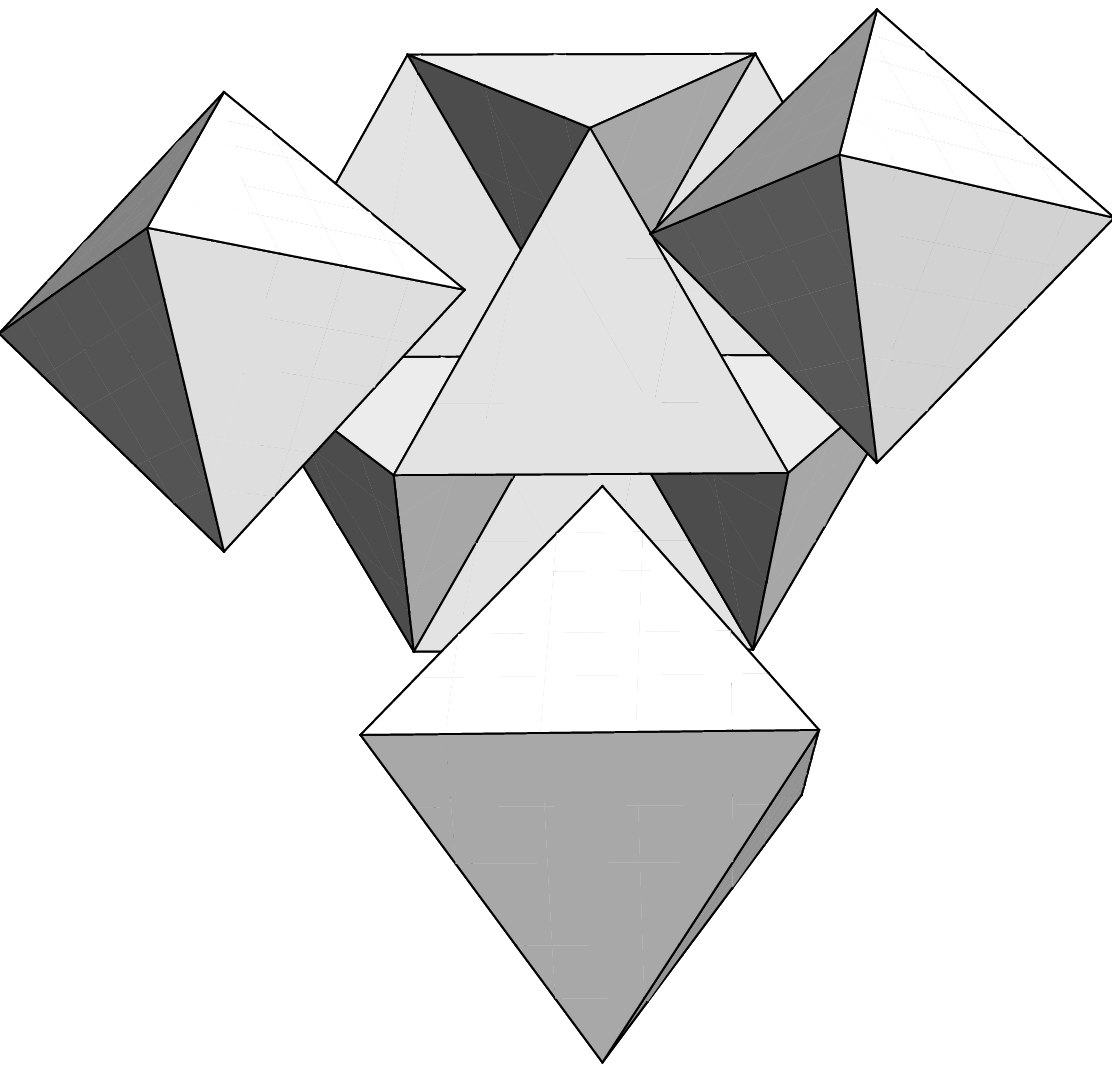}}
\caption{\small{All the octahedron's edges are divided into equal halves. Connecting the points results in $4\times 8=32$
triangles. On (a) the white triangles are the faces of the
second-generation tetrahedra.  Each face contributes one
tetrahedron, which are shown in (b), inside the outline of the
parent octahedron. Each of the octahedron's vertices contributes a
child octahedron. Three of them are shown on (c), where they
complete the eight tetrahedron complex to the parent octahedron.
 of the ch}}
\end{figure}
This similarity can be established either by elementary geometry,
using Fig.~3 as an aid, or by finding out the explicit
transformation law. As an example, consider the secondary
tetrahedron $(bt_2t_3t_4)$. Vertices of the initial tetrahedron are
mapped into its vertices according to
\be
v\mapsto b+\frac{1}{4}v_1-\frac{1}{4}v.
\ee
Indeed, $v_1\mapsto b\equiv b+\frac{1}{4}v_1-\frac{1}{4}v_1$, and
$v_i\mapsto t_i$.

If at some stage the triangulation consists of $T$ tetrahedra and
$O$ octahedra, then one refitment step results in
\be
T\mapsto 4T+8O, \qquad O\mapsto T+6O.
\ee
Consequently, after $n$ subdivisions
\begin{align}
T_n&=\frac{1}{3}\left(2^{3n}+2^{n+1}\right)\sim \frac{2^{3n}}{3},
\\
O_n&=\frac{1}{3}\left(2^{3n-1}-2^{n-1}\right)\sim\frac{2^{3n}}{6}.
\end{align}
Hence the volume fraction of the tetrahedra that are similar to the
initial one asymptotically reaches $\frac{1}{3}$. Since the final
iteration is followed by a subdivision of the octahedra into four
tetrahedra, the $n$ step iterative procedure gives us $T_n$
tetrahedra that are similar to the original one (and their volume
fraction asymptotically reaches $\frac{1}{3}$), and $T'_n=4O_n$
tetrahedra of other classes.

\begin{figure}[floatfix]
    \begin{minipage}{\columnwidth}
    \begin{center}
        \resizebox{0.7\columnwidth}{!}{\includegraphics{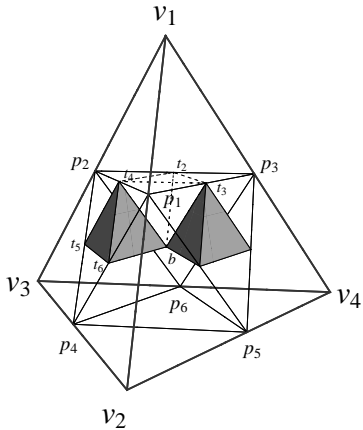}}
    \end{center}
    \end{minipage}
    \caption{\small{
 The barycenter $b$ is a common vertex of all
the second generation tetrahedra. Two of them are shown as solid
bodies, and one as a dotted outline. For a general initial
tetrahedron the  octahedron's diagonals $(p_1p_6)$, $(p_2p_5)$, and
$(p_3p_4)$ do not intersect at the
barycenter.}}\label{choiplot}\vspace{-5mm}
\end{figure}
 Assume that the number of steps is such that the surface areas of
the small tetrahedra $T$ (and two out of four faces of the
tetrahedra $T'$) still satisfy $2A_n=J_n\approx j_*\gg 1$. Hence
$2A=2^{2n}j_*\equiv Kj_*$. To establish our point in the simplest
possible way we focus only on the tetrahedra $T$.  The
dimensionality of a (SU(2) invariant) Hilbert space that is
associated with a single such tetrahedron is $d_*=2j_*$, and their
total number is $K^{3/2}/3$. The total Hilbert space dimension is
\be
d(K)=(2j_*)^{\frac{1}{3}K^{3/2}}=(2j_*)^{\frac{1}{3}\left(\f{2A}{j_*}\right)^{3/2}}.
\ee
As a result,  if the shape is encoded in the information-theoretical
optimal way, the shape uncertainty decreases super-exponentially, as
\be
\Delta(z_{12})\Delta(z_{23})\propto \f{1}{d(K)}.
\ee
This result ignores the expectation of the volume operator, and its
usefulness is mainly in setting the upper limit on the convergence
to classicality.

We can use the same construction to show a modest improvement even
when the total measured state is given by
\be
|\Psi\9=\otimes_a^{\frac{1}{3}K^{3/2}}|\psi\9_a,
\ee
and the state of each $|\psi\9$ is given by Eq.~(\ref{state}). If
the  (commuting!) $J_{12}$ and $J_{23}$ are estimated for all the
small tetrahedra independently by separately applying a POVM of
Sec.~III to each tetrahedron,   then the statistical averaging over
the entire sample leads to
\be
\Delta(z_{12})\Delta(z_{23})\simeq\f{3c}{2j_*}\f{1}{K^{3/2}},
\ee
where a constant $c$ is  determined by the asymptotics of a single
tetrahedron. As a result, the uncertainty  goes to zero faster than
the uncertainty of Eq.~(\ref{upper})



\section{Conclusions}
We constructed a SU(2)-invariant positive operator valued measure
that simultaneously extracts two classical parameters that are
associated with non-commutative observables. It provides a new
method to use SU(2) coherent states to build gauge-invariant
objects, thus complementing the analysis of
\cite{etsi}. Mapping the semiclassical problem into a
quantum-informational task, we showed that for a single tetrahedron
the fuzziness of geometry is reduced only as  an inverse of the
area. However, a judicious choice of more complicated states
speeds-up the convergence. It still remains to be seen weather an
exponential convergence to the classical limit is possible.

\acknowledgments
Discussions with Etera Livine and Jimmy Ryan are gratefully
acknowledged.

\section*{Appendix A}
In elementary quantum measurement theory, a test performed on a
finite-dimensional quantum system is represented by a complete set
of orthogonal projection operators $\hat{P}_m$, where the label $m$
takes at most $d$ different values (d is the dimensionality of the
Hilbert space. The probability of obtaining outcome $m$ of that
test, following the preparation of a quantum ensemble in a state
$\rho$, is
$$p_m(\rho)=\tr \rho \hat{P}_m.$$
 In the
infinite-dimensional case such as, e.g., the space of a single
one-dimensional non-relativistic particle, the measurement outcomes
are associated with spectral decomposition of self-adjoint
operators. For example a projection-valued measure for finding a
particle in the segment $(a,b)\in\mathbbm{R}$  is written in the
improper position basis $|x\9$ as
\be
\hat{P}((a,b))=\int_a^bdx|x\9\6x|.
\ee

 It is well known that this framework is not suitable for description of joint measurements of non-commuting
observables, such as non-relativistic position and momentum. Tests
of this type are not  optimal for many quantum-informational tasks.
Moreover,  measurement of certain classical quantities (phase, time,
relativistic spacetime localization) cannot be described at all in
this language.

Those difficulties are overcome with the help of generalized
measurements which are described by positive operator-valued
measures (POVM). Those are essentially non-orthogonal decompositions
of identity by positive operators. Unlike the standard (von Neumann,
or projective) measurement descriptions they do not provide a
spectral decomposition of some self-adjoint ``observable", while the
rest of the rules are kept intact. E.g.,a finite set of outcomes
$\mu$ is associated with  positive operators $\hat{E}_\mu$ that
satisfy
\be
p_\mu(\rho)=\tr(\rho \hat{E}_\mu), \qquad \sum_\mu \hat{E}_\mu=\1,
\ee
but there is no requirement of $\hat{E}_\mu
\hat{E}_\nu=\delta_{\mu\nu}E_\nu$. In a finite-dimensional setting it
allows to consider an arbitrary number of the measurement outcomes.
Covariance considerations play an important role in constructing
POVMs and finding the optimal protocols for particular
quantum-informational tasks. Their theory is well-developed and is
one of the cornerstones of quantum information theory.

We have to keep in mind two related features. First, compared to a
corresponding projective measurement, a generalized measurement is
less sharp. For example, a Heisenberg uncertainty relation reads
$\Delta q\Delta p\geq
\hbar/2$, where  statistics is taken over an ensemble of identically
prepared systems, with position and momentum measured separately on
the half of systems each. The optimal POVM for a joint position and
momentum measurement results in $\Delta q\Delta p\geq
\hbar$
Second, statistical moments in a projective measurement
\be
\hm^{(1)}=\sum_mx_m\hat{P}_m, \qquad \hm^{(2)}=\sum_m
x_m^2\hat{P}_m, \qquad \ldots
\ee
have simple relations with each other, such as
\be
\hm^{(2)}=(\hm^{(1)})^2.
\ee
This is not true for the statistical moments derived from a POVM.

A detailed exposition of POVM theory can be found, e.g., in
\cite{hol, bgl, rmp}.

\section*{Appendix B}

In this Appendix we gather the explicit formulas for the matrix
elements of the first two statistical operators of the POVM of
Sec.~\ref{joint}. After the angle integration the first moment
operator $\hm^{\!(1)}$ of Eq.~(\ref{firstm}) becomes
\begin{widetext}
\begin{align}
& \hm^{\!(1)}=\sum_{m_1,m_2} m_1m_2|m_1m_2\9\6m_1m_2|+ \nn\\
&\f{1}{2}\left(\sum_{m_1=-j}^{j-1}\sum_{m_2=-j+1}^jf(m_1,m_2)|m_1m_2\9\6m_1+1m_2-1|+
\sum_{m_1=-j+1}^j\sum_{m_2=-j}^{j-1}f(m_2,m_1)|m_1m_2\9\6m_1-1m_2+1|\right),
\end{align}
where
\be
f(m_1,m_2)=\sqrt{(j-m_1)(j+m_2)(j+m_1+1)(j-m_2+1)}.
\ee
The second moment operator has a tri-diagonal form,
\be
\hm^{\!(2)}=A+B+C
\ee
where

\be
A=\f{1}{(2j+3)^2}\sum_{m_1m_2}\left((j+2m_1^2+1)(j+2m_2^2+1)+2[(j+1)^2-m_1^2][(j+1)^2-m_2^2]\right)(|m_1m_2\9\6m_1m_2|
\ee
\be
B=\sum_{m_1m_2}B(m_1,m_2)|m_1m_2\9\6m_1+1m_2-1|+B({m_2,m_1})|m_1m_2\9\6m_1-1m_2+1|,
\ee
where
\be
B(m_1,m_2)=(2m_1+1)(2m_2-1)\sqrt{(j-m_1)(j+m_2)(j+m_1)(j-m_2+1)},
\ee
and, finally,
\be
C=\sum_{m_1m_2}C(m_1,m_2)|m_1m_2\9\6m_1+2m_2-2|+C({m_2,m_1})|m_1m_2\9\6m_1-2m_2+2|,
\ee
where
\be
C(m_1,m_2)=\sqrt{(j+m_2)(j+m_2-1)(j-m_2+1)(j-m_2+2)(j-m_1)(j-m_1-1)(j+m_1+1)(j+m_1+m_2)}
\ee
Expressions for these operators in $|l,m\9$ basis with the help of
usual SU(2) recoupling relations. Since the moment operators result
from a POVM,
\be
\hm^{(2)}\neq (\hm^{(1)})^2.
\ee
\end{widetext}

\end{document}